# A Reference Framework for Variability Management of Software Product Lines

Saiqa Aleem[1*], Luiz Fernando Capretz[2], & Faheem Ahmed[1]

[1] Department of Engineering, Thompson Rivers University, Kamloops, British Columbia, Canada, V2C 0C8

[2] Electrical & Computer Engineering Department, Western University, London ON, Canada, N6A 5B9

Correspondence: Saiqa Aleem Department of Engineering, Thompson Rivers University, Kamloops, British Columbia, V2C 0C8, Canada.

**Abstract**

Variability management (VM) in software product line engineering (SPLE) is introduced as an abstraction that enables the reuse and customization of assets. VM is a complex task involving the identification, representation, and instantiation of variability for specific products, as well as the evolution of variability itself. This work presents a comparison and contrast between existing VM approaches using "qualitative meta-synthesis" to determine the underlying perspectives, metaphors, and concepts of existing methods. A common frame of reference for the VM was proposed as the result of this analysis. Putting metaphors in the context of the dimensions in which variability occurs and identifying its key concepts provides a better understanding of its management and enables several analyses and evaluation opportunities. Finally, the proposed framework was evaluated using a qualitative study approach. The results of the evaluation phase suggest that the organizations in practice only focus on one dimension. The presented frame of reference will help the organization to cover this gap in practice.

## 1. Introduction

Since the 1960s, the main concern in software engineering has been the creation of software products in a shorter time to market with low cost and high quality. By addressing these concerns, software product line engineering (SPLE) has attracted substantial attention in recent years because it promises the construction of high-quality software products at a lower cost in less time (Pohl et al., 2005); Moon et al. (2005)) by proactive use. Furthermore, SPLE supports the systematic development of a group mileage method for identifying variation and mapping it to other phases, that is, domain engineering and application engineering. Moreover, the existing approaches in VM literature appear to be incongruent with each other because they take different perspectives and use somewhat different metaphors with limited scope (Chen et al. 2009). These problems highlight the need for a common frame of reference for VM adoption and implementation in SPLE because organizations are exposed to several approaches. A common frame of reference would help developers understand and implement VM effectively in different views of domain engineering and application engineering (SPLE).

This study aims to provide a common frame of reference for VM in SPLE by using "qualitative meta-synthesis" (Walsh & Downe, 2005). The qualitative meta-analysis approach used comparison and contrast techniques. The comparison and contrast technique between existing VM approaches helps determine the underlying perspectives, metaphors, and concepts of existing methods. The study adopted a qualitative meta-synthesis approach and followed the research methodology explained by Walsh & Downe (2005).

*1.1 Related Work*

Many software product-line design methods and methodologies, including but not limited to FODA (Kang et al., 1990), FORM (Kang et al., 1989), FAST (Ardis et al., 2000), SPLIT (Coriat et al., 2000), KobrA (Atkinson et al., 2000), and QADA (2019) have been proposed to manage variabilities in software product lines over the past two decades with applications in different industrial contexts. Recently, Itzik et al. (2016) introduced an approach called semantic and ontological variability analysis (SOVA) to analyze the variability among different software products based on their textual requirements. In SOVA, two perspectives are discussed: structural and functional, and its outcomes are feature diagrams organized according to selected perspectives. FeatureIDE (Meinicke et al., 2016) is an Eclipse-based tool that supports the handling of multiple challenges with preprocessors and is used in the entire life cycle of software product life development.

Many studies have analyzed variability management techniques from a different perspective. Matinlassi (2004) provided a detailed comparison of software product line architecture design methods: COPA, FAST, FORM, KobrA, and QADA by their context, user, structure, and validation. Chen et al. (2009; 2011) carried out a systematic literature review of VM approaches in a software product line from 1990 to 2007. They reported that most VM approaches were neither evaluated thoroughly using scientific techniques nor the trends of scientific evaluation of VM approaches appeared to be improving. They also highlighted areas that require improvement and proposed that these approaches can be constructive if applied appropriately. Sinnema et al. (2007) provided a detailed classification of variability modelling techniques based on their similarities. They also highlighted the differences, scope, size, and application domains of product families.

Similarly, Metzger & Pohl (2014) briefly summarized the significant research achievements and structured their research summaries along with a standardized software product line framework. They found significant contributions in the areas of variability modelling and informally verified product line artifacts. They also highlighted trends that will influence software product line engineering research in the coming years, which shows research opportunities at the intersection between software product line engineering and service-oriented computing, cloud computing, big data analytics, autonomic computing, and adaptive systems.

Moreover, Arrieta et al. (2015) analyzed variability challenges in a cyber-physical system with a resulting taxonomy that is useful for future developers of CPS product lines. A systematic literature review by Galster et al. (2014) for variability highlighted that throughout the software engineering phases, only testing is a phase where variability has not been addressed sufficiently. They reported that for variability management, software quality attributes have not gained much consideration. The proposed dimensions of variability indicate many opportunities for future research. Reinhartz-Berger et al.(2017) investigated the comprehensibility of the two methods of variability organization into models. They concluded that when creating a model from natural language description, the modelling style is determined by prior exposure to the modelling style and the degree of dependency among elements. The comprehension of the variability model also depends on the degree of dependence.

Empirical studies on variability management comprise the work of Chen & Babar (2010), who identified the issues and challenges of VM faced by industry practitioners. Some of these challenges were later confirmed by Berger et al. (2013). Berger et al. (2013) tried to improve the empirical understanding of variability modelling practices in industrial software product lines by analyzing the survey results of industrial practitioners. They found that most models used in industry have variability units ranging from 50 to 10,000 and a high heterogeneity of notations and tools. Czarnecki et al. (2012) compared FM and decision modelling approaches for variability. They structured the research conducted in the field of variability modelling and evaluated possible synergies. The comparison consists of many aspects of FM and DM, including historical origins and rationale, syntactic and semantic richness, tool support, and identifying commonalities and differences. The organizational factors and business factors of successful software produtct lines were investigated by Ahmed et. al. (2007) and Ahmed and Capretz (2007) respectively; these two studies led to the creation of a business maturity model of software product line engineering (Ahmed and Capretz, 2011). Bosch et al. (2015) presented trends in software variability in the technical practice area.

To build systems that are more context-aware, post-deployment reconfigurable and runtime adaptive dynamic software product line (DSPL) engineering exploit the knowledge acquired in SPLE (Melo et al., 2007;2013). Da Silva et al. (2016) provided a systematic review of the literature. They identified the assets, activities, tools, and approaches used in requirements engineering and variability management in dynamic software product lines (domain engineering). The activities are focused on DSPL modelling and specification. They mentioned that traditional approaches such as UML diagrams could be used to document the domain requirements as well as the feature model, which can also be used to represent the domain variability. Guedes et al. (2015) conducted a systematic mapping to discover how variability is modelled in DSPL approaches and which information is used to guide variability binding at runtime. They synthesized the results of the systematic mapping, which can be used to identify trends and gaps for research on the variability management of DSPL. They found that various notations were used and discovered that two-thirds of their selected studies used feature models to capture variability. The proposed framework of Bashari et al. (2017) is designed by defining a set of dimensions that answer questions about how runtime adaptation can be realized using DSPL engineering approaches. For the organization of dimensions, their framework conceptualizes DSPL adaptation management as a MAPE-K loop.

The related work on variability management mostly includes quantitative-based research. Moreover, the approaches mentioned above only cover a single aspect of VM, such as analysis, implementation, adoption, domain engineering, or application engineering phase. There is a lack of a common frame of reference in SPLE for VM adoption and implementation. In this study, we presented the VM common frame of reference that covers the domain and application engineering phases as well as business and technical aspects. We used the qualitative meta-synthesis method to determine the underlying perspectives, metaphors, and concepts of existing VM approaches and proposed a common frame of reference that would help developers understand and implement VM effectively in different views of SPLE.

The rest of the paper is structured as follows: Section 2 (Research motivation and methodology) introduces the qualitative meta-synthesis procedure used in this study. Section 3 (Results and analysis stage of meta-synthesis techniques) reviews and analyzes eleven VM approaches found in the literature between 2000 and 2019. The existing VM approaches are compared and contrasted with each other, revealing the perspectives, scope, and metaphors used by each.

Section 4 defines the final VM framework. Section 5 (a common framework for variability management) proposes a common frame of reference for describing and implementing VM based on the analysis presented in the results section, that is, Section 3. Section 5 presents the evaluation of the proposed framework. Section 6 (Conclusions and future research directions) presents the concluding comments and directions for future research.

**2. Research Methodology and Motivation**

*2.1 Research Motivation*

This research aims to reveal the underlying concepts and metaphors present in various VM approaches and to theorize and build an underlying higher-level VM model to achieve a better understanding by developers. The main research motivation behind this research is that VM is one of the fundamental activities of SPLE and involves the explicit representation of the variation in software product artifacts, management of dependencies among variants, and support for instantiations of the product line throughout the SPLE lifecycle. In product line engineering, efficiently realizing and managing variability is a key challenge (Myllärniemi et al., 2016). Therefore, it is clear that VM involves extremely challenging and complex tasks, which must be supported by a common frame of reference that clarifies the decision model for variations and endorses the use of effective techniques and tools. Several approaches have proposed solutions to these challenges for nearly 20 years.

The goal of this research is to synthesize the VM approaches proposed by researchers during the last two decades and develop a common frame of reference for VM description and implementation throughout the SPLE lifecycle. The proposed approaches were identified through a literature review. From the proposed approaches, various metaphors and perspectives observed were contrasted, compared, and synthesized. It is noteworthy that the research approaches found in the literature review are based on quantitative rather than qualitative studies. Qualitative studies provide depth and details for the analysis, rather than ranks based on quantitative analysis. They also created openness for further analyses. Therefore, a qualitative meta-synthesis approach appears to be a suitable technique for exposing the underlying frame of reference for VM description and implementation. Specifically, the qualitative meta-synthesis approach presented by Walsh and Downs 2005 was chosen for this study. Walsh and Downe (2005) developed and proposed a six-step approach for qualitative meta-synthesis through an extensive literature review using different phrases such as "meta-analysis", "meta-synthesis", and follow-up "berry picking" (Bates, 1989) procedures. The six steps of the qualitative meta-synthesis approach include: (a) framing the meta-synthesis exercise, (b) locating relevant papers, (c) deciding what to include, (d) appraising the studies, (e) performing analytical techniques such as comparing, contrasting, and reciprocal translation, and (f) synthesizing the translations.

*2.2 Research Methodology: Qualitative Meta-Synthesis*

In this study, the "qualitative meta-synthesis" research methodology was used (Walsh and Downe, 2005). This is recognized as an exploratory research approach that is used to extract or build a common frame of reference from the analysis of research results. This approach helps to create a staged framework. In the healthcare research domain, the phrase qualitative meta-synthesis was first coined by Stern and Harris (1985) in the context of a systematic review of qualitative studies, but it must be distinguished from the concept of a systematic review. A systematic review can be performed by following a series of well-defined steps. Statistical analysis was performed on a pool of research studies, enabling a robust comparison among quantitative studies. This type of statistical analysis can also be called a meta-analysis. Studies based on qualitative meta-synthesis can provide more in-depth analysis and investigation. The objective of qualitative meta-synthesis is to develop a model or exploratory theory that can explain the results of a group of similar qualitative studies (Dixon-woods et al. 2007; Finlayson and Dixon, 2009; Humphrey et al. 2007; Jensen and Allen 2006). Such an aggregation of qualitative studies has been described by Zimmer (2006): "through a process of translation and synthesis, identification of consensus, hypothesis development, and investigation of contradictions in patterns of experience across studies make theorizing at higher levels possible' (p. 1). This translation and synthesis seek not only to retain the distinctive features of individual studies but also to reveal their differences, thereby facilitating understanding by both researchers and readers of how various research results are related to each other.

In this study, the qualitative meta-synthesis procedure is outlined in Figure 1, followed by the details of each phase in the following sub-section.

1) Design Framing a Qualitative Meta-synthesis:

The first step of a qualitative meta-synthesis approach involves an appropriate research question. First, a research question must always be proposed. Selecting a topic for qualitative meta-synthesis is critically important because many factors such as research gaps, research impact, and individual or community interests contribute to shaping this research question.

2) Locating Relevant Studies and Deciding What to Include:

This step involves searching for relevant studies in the available databases, which can be performed in two phases. In the first phase, an initial search can be performed based on the identified keywords and the research question that was formed in the previous step. An initial selection can be made based on the suitability criteria for the research topic. The second phase of screening was based on the Bates (1989) "berry picking" approach. The main purpose of this approach is to search for related "approaches used for the study", and a citation analysis was performed by searching for "mainstream approaches" and following the chain of citations. This approach will help to include any important studies that were missed in the previous step.

3) Appraisal Studies:

This phase involves screening studies with low quality. Walsh and Downe's (2005) approach was selected for this study, except for the appraisal phase.

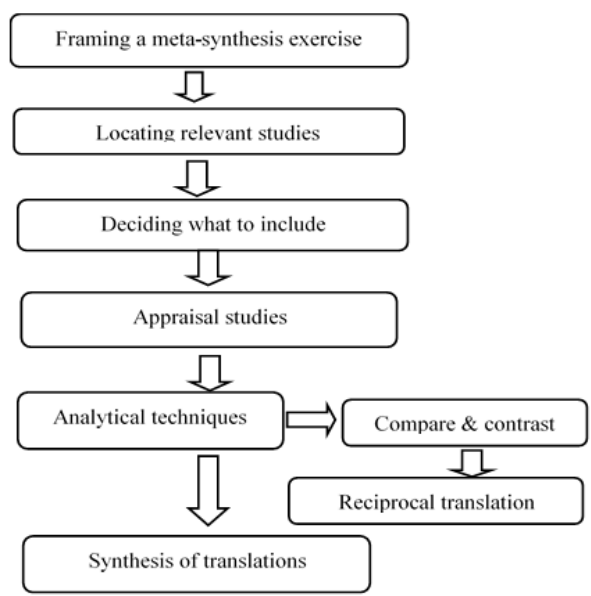

Figure 1. Qualitative meta-synthesis procedure (Moon et al. 2005)

4) Analytical techniques

This step involves the "compare and contrast phase" and reciprocal translation of the approaches of the identified studies. In the comparison and contrast phases, the researcher determines how frameworks/approaches overlap or differ in the selected studies. The reciprocal translation step involves exploring the metaphors involved in the identified concepts and themes for reciprocal translation that the various approaches could use. Themes and concepts are linked to the identification of overarching metaphors. Meaningful dimensions were identified across the different phases during this process.

5) Synthesis of Translations:

Progressively, clusters of metaphors, themes, and concepts can be refined and can emerge as substantive theories (Sherwood, 1997; Strauss & Corbin, 1998). In this step, translation synthesis addresses the contradictions and overlaps in the concepts identified in the reciprocal translation process.

*2.3 Application of qualitative Meta-synthesis*

In the following section, the phases of the meta-synthesis procedure are described.

1) Framing a qualitative meta-synthesis:

The research question covered in this study includes the interrogation of underlying metaphors and themes in various VM approaches found in the literature and the development of a common frame of reference that developers can use to design and implement VM initiatives. First, we defined the general question of the study as follows:

General Research Question: Can we develop a common framework for VM description or its implementation for developers?

To answer the general question, we defined the sub-questions which are listed below:

Research Question (i): What are the themes and metaphors contained in various existing VM approaches?

Research Question (ii): What is the common frame of reference for VM implementation?

2) Locating Relevant Studies and Deciding What to Include

First, a conventional electronic database search was undertaken using various combinations of terms such as variability, approaches, management, software products, implementation, tools, variants, modelling, and model against major electronic databases, including IEEE Computer Society, ACM Communications, Google Scholar, Elsevier Science Direct, Wiley Online Library, and Springer. This procedure, as explained earlier, consists of two phases. In the first phase, an initial search was performed, 334 articles were retrieved based on the identified keywords, and a research question was formed. An initial selection was made based on the following suitability criteria for the research topic:

- The studies should be about VM approaches, their description, and implementation.
- The full article text should be available.
- If any article identifies any framework or approach for implementing VM in SPLE, that article is included.
- Studies that described the modelling of VM in SPLE were included.
- Studies that described the importance of VM and discussed its issues were included.
- Analyses of case studies for VM were included.

Some articles were excluded because they were not directly related to VM for SPLE based on the following exclusion criteria:

- Articles published on company Web sites were not included.
- Articles irrelevant to the research questions were not included.
- Articles that did not describe VM implementations were not included.
- Articles that did not identify VM approaches in SPLE were not included.

Accordingly, 233 research papers were selected. In the second phase, 101 articles that were directly related to VM approaches based on the inclusion and exclusion criteria shown in Figure 2 were selected. The second phase of screening was based on the Bates (1989) "berry picking" approach. The main purpose of this approach was to search for related "approaches used for VM implementation by developers", and a citation analysis was performed searching for "mainstream approaches of VM for SPLE and its implementation" and following the chain of citations. As a result, some researchers have used previously developed techniques to support their analyses or arguments. At this stage, various relevant articles from the literature or elsewhere, such as industry reports, book chapters, and white papers from international organizations, were identified and located from journal citations through recursive research. Using Google and Google Scholar Web Search, this recursive research was augmented. As a result, 17 VM approaches were identified, covering different aspects of SPLE.

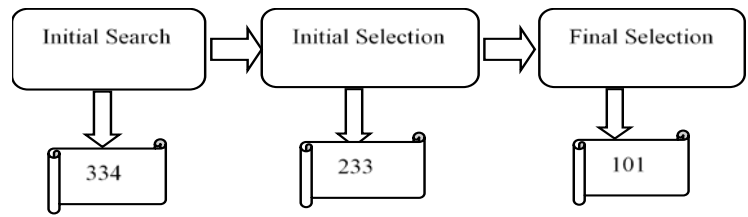

Figure 2. Study selection process

3) Appraisal Studies

This approach is suitable only in areas such as education, nursing, and healthcare, where many studies have been conducted. In the final selection of articles on VM for SPLE, relatively few studies remained, and because the objective of this research was to highlight the metaphors and concepts underlying these approaches, the rigorous relevance of the approach was more critical than the implementation process itself. Therefore, judgments were not made about the quality of the articles or the approaches, and any approach suggesting the modelling or description of VM at any point in time in SPLE or its implementation was included in this analysis. In other words, the appraisal phase was conducted under relaxed criteria.

4) Analytical techniques

The identified VM approaches for SPLE were compared and contrasted with each other from various perspectives. Practically, this process involved in-depth reading of each article and report. Understanding the author's usage of key ideas, metaphors, phrases, relations, and concepts generated a grid that linked concepts to themes by tabulating them against each other. This is an interpretive and descriptive process. As Jensen and Allen (2006) explained, "this is a two-part process. The first part accurately captures concepts, while the second is a dialectic relating of studies to each other through the juxtaposition of the concepts identified in the process." By using this juxtaposition of concepts, the identified terms revealed degrees of heterogeneity or homogeneity among these approaches, as can be seen in new concepts for a common frame of reference for VM. The results of this phase are presented in Table 3, Appendix I, and Figure 4 and are discussed in detail in Section 3. In short, this research is based on a qualitative meta-synthesis research approach that combines results from various qualitative approaches within a specific domain. The domain of interest for this study includes VM approaches, and the study follows a detailed procedure as suggested by Walsh and Downe (2005) and also adopted by Lee (2010) to answer this research question and its sub-questions. However, this approach cannot be considered as a literature review or an analysis of primary data, but rather as a qualitative study aggregating results interpreted by researchers in a collection of VM approaches found in the literature. The outcomes of this phase are presented in Table 1 in Appendix I. Table 1 includes columns that list each identified approach's underlying concepts, modelling techniques that they cover, the scope of each approach, each approach addressed issues and the limitations of each approach. Table 1 is discussed in detail in section 4.1.

5) Reciprocal translation

This process was reciprocal in the sense that, among the VM approaches, the comparison and creation of concepts and metaphors were performed repeatedly. In particular, when concepts are homogeneous in terms of usage and definitions, this process is relatively straightforward. At other times, the same concepts may stand in opposition to each other. Noblit and Hare (1988) described this as "refutation translation" because concepts may overlap without being sufficiently replaceable. This translation may be a sign of the existence of a new category that was not revealed in the first step. The results of this reciprocal translation are shown in Table 2 in Appendix I. Table 2 has three columns: the first column lists the approach, the second column lists the SPL stage described in the approach, and the third column provides the mapping in comparison with SPL stages for variability management. The outcome is discussed in section 4.2 in detail.

6) Synthesis of Translations

The last phase of the meta-synthesis approach involved the synthesis of translated and juxtaposed metaphors and concepts in elucidating exploratory theories, underlying dimensions, and new concepts for a common frame of reference for VM. The results of this phase are presented in Table 3, Appendix I, and Figure 4 and are discussed in detail in Section 4.3.

In short, this research is based on a qualitative meta-synthesis research approach that combines results from various qualitative approaches within a specific domain. The domain of interest for this study includes VM approaches, and the study follows a detailed procedure as suggested by Walsh and Downe (2005) and also adopted by Lee (2010) to answer this research question and its sub-questions. However, this approach cannot be considered as a literature review or an analysis of primary data, but rather as a qualitative study aggregating results interpreted by researchers in a collection of VM approaches found in the literature.

## 3. Results and Analysis of Meta-Synthesis

This section elaborates on the results of the analytical analysis stage of the meta-synthesis technique. From this stage, twelve approaches were identified after the recursive and iterative literature search and analysis of citations described in the section on the detailed qualitative meta-synthesis process. A subsequent web search was performed after studying the documentation and original reports from companies and consulting firms and confirmed the use and existence of these approaches (FODA (Kang et al. 1990), Koalish (Asikainen et al. 2004), KobrA (Atkinson et al. 2000), Kumbang (Asikainen et al., 2007), COVAMOF (Junior et al. 2005), and others). Most of these approaches include custom tools to support product line activities as part of their methods (Chimalakonda et al., 2016).

Most VM methods have been developed by individual researchers and have been described in the academic literature. It seems that some of these approaches were based on previously developed approaches or combinations of more than one VM approach, whereas others were disconnected in terms of their specific details as well as their general perspective. The approaches studied are quite diverse in terms of their goals, design philosophy approaches, and variability modelling methods. Because of the diversity of the reported approaches, it would be difficult to classify them, although some researchers have tried. This study attempted to describe the approaches based on the underlying classification strategy. The outcome of the analytical technique phase is mentioned in Table 1, Appendix I. Clearly, several approaches cover only one or two phases of the SPLE life cycle. Only a few approaches have been found that attempts to cover the full SPLE lifecycle; these are presented in Table 2 in Appendix I.

*3.1 Result of Analytical Technique Phase*

A brief description of each identified approach as an outcome of the analytical technique phase is provided below:

**Feature-oriented domain analysis (FODA)** based VM approaches: The FODA (Kang et al. 1990) method developed by the Software Engineering Institute (SEI) uses features to characterize a domain. The features of a product can be services, characteristics, or technologies in a particular product line. This approach uses features to control variability in both problem and solution spaces and introduces the basic concept of feature modelling. From 1998 to 2008, researchers proposed and implemented numerous extensions and enhancements to the original FODA model (Bosch et al. 2015). The feature-oriented reuse method (FORM) (Knag et al. 1989) is an extension of the FODA approach that supports VM. It identifies the commonalities among applications in a particular domain and constructs a feature model during the analysis. This featured model captures commonalities as an AND/OR graph, mandatory features can be identified by AND and OR nodes, and alternative features selectable for different applications can be specified. Many approaches have been proposed to use the same feature modelling concept or extend it. A tool for VM in the requirements phase was developed by RequiLine (von der Massen et al., 2004). This approach proposed a model that reused features as an extension of a feature model. These features might be useful or mandatory in one domain, but optional in another. Ferber et al. (2002) used separate views to represent feature dependencies and interactions. The same approach was extended by Ye and Liu (2005) by expanding the meaning of the views.  Many researchers have studied state-of-the-art feature model analysis (Benavides et al., 2010; Lesta et al., 2015; Thum et al., 2014). Sree-Kumar et al. (2016) reviewed the state-of-the-art analysis of FMs using alloy (a popular framework used for feature model analysis) using the list of analysis operations as proposed by Benavides et al. (2010) as a reference. Identification of features, their representation in the feature model, and then variability design from architecture to component implementation are stages of FODA-based approaches. Approximately 13 approaches are based on FODA, and these approaches help in analyzing domain requirement commonalities and performing variability modelling or analysis. Only FORM provides comprehensive coverage of the SPLE lifecycle, excluding the testing phase.

**Koala-Based VM Approaches:** Koala (Asikainen et al., 2004) was developed by Philips Consumer Electronics for the development of embedded software in consumer electronic devices. The modelling elements of Koala contain components that have interfaces, functions corresponding to function signatures, and configurations that are not part of components and have no interfaces, as well as bindings between interfaces. These elements are used to specify the logical structure of a software system. Some of these VM approaches were inspired by Koala (Asikainen et al., 2004). VM is performed at compile time using Koala-based approaches. The method proposed by van der Hoek (2004) differs from Koala because it supports variability management at any point in the software lifecycle. Another extension of the Koala approach is Koalish (Asikainen et al., 2004), a component model, and an architectural description language for explicit VM mechanisms that are used to capture alternative or optional components. For contained components, a set of possible types and cardinalities can be defined to enable variability tracing among them. Constraints can be used in this approach to restricting the set of valid individual systems. Koalish is based on the product configuration domain. Kumbang (Asikainen et al., 2007) is a combination of Koalish and feature modelling concepts. It models SPLE from both features and architectural perspectives. This approach was also introduced by the common author of Koala and Koalish. It is based on three layers of abstraction. The meta-layer has the highest level of abstraction and is made up of modelling concepts or metaclasses. The model layer is the next layer consisting of the Kumbang models; it contains entities and classes that are instances of metaclasses. The final layer is the instance layer and consists of instances of model layer classes. Variability is represented in a Kumbang model that contains a set of Kumbang types and is the description of its instances. Kumbang synthesizes existing methods of variability modelling and provides a basis for the development of VM modelling and tool support from requirements specification to architecture phases (Asikainen et al., 2007), as well as enabling SPL configuration according to specific customer requirements.

**Component-based product-line engineering VM approaches:** The KobrA approach serves as the foundation for SPLE and is based on the concept of component-based product-line engineering (Asikainen, 2002). It uses a decision model to represent variability identification and design (Asikainen et al., 2001). The decision model consists of variability IDs, variation points, effect sets, and a resolution set for variability. A questionnaire was used for requirements identification, after which variability was derived. The derived variability was represented in terms of the variation points with a resolution setting. This approach does not explicitly present the variability types and scope. Another component-based variability approach was proposed by Bachmann and Bass (2001) for an SPLE architecture. In this architecture, different variation points were identified as data, functions, control flows, quality goals, technologies, and the environment. Variability features can be an alternative, a set of alternatives, or optional. This approach does not address the appropriate scope or level of detail for variability types. The approach proposed by Muthig and Atkinson (2002) is also based on a decision model approach for the architectural design phase. It introduces three types of variation points. However, this approach is limited to a known range of variability. Lau et al. (2014) surveyed and studied existing component-based software engineering approaches and their corresponding component models. They also defined a new component model and the taxonomy of the existing component model. Few studies using the component-based development approach provide a classification of variability in terms of its scope (Lee et al., 1999; Sharp, 2000; Becker et al. 2002).

**UML-Based VM Approaches:** The UML process-based approach was introduced for variability management (Junior et al., 2005). It provides identification, representation, and delimitation of variability, as well as the identification of mechanisms for variability implementation. A systematic review of the evaluation of variability management (Chen & Babar, 2011) showed that a large majority of VM approaches are based on feature modelling and/or UML-based techniques. The UML-based process supports the SPL lifecycle, variability tracing, and analysis of specific product configurations. Generally, variability refers to the variable aspects of the SPL products. Term variation and variants are also used to describe variability (Pohl et al. 2005). The variability management process runs in parallel with the development of core assets because it is an iterative and incremental process. In the core asset development process, the variability management process is executed after each activity implementation. The number of variabilities can be expected to increase as activities are executed, and updates of variability are allowed from any activity in the process. The proposed process is constructed as follows: a) the variability tracing definition stage accepts inputs in the form of a use case and a feature model and generates output in the form of a variability tracing model; b) the variability identification stage accepts the use case, the feature model, the static type, and the component model as inputs and generates the identified variabilities along with the same artifacts as outputs; c) variability delimitation accepts the same inputs as the variability identification stage and generates outputs in the form of the same artifacts with the variabilities limited; and d) identification of mechanisms for variability implementation accepts the static type and the component model as inputs and generates output in the form of a variability implementation model. The evaluation method was based on case studies, and the results showed that it is beneficial to establish well-defined and controlled variability management for core asset development as the main activity. This study also proposed a metadata model that constitutes the basis for tool design to support variability management. The proposed approach is limited to domain engineering.

**Systematic process-support-based VM Approach:** Family-Oriented Abstraction, Specification, and Translation (FAST) (Ardis et al. 2000) use a systematic process for commonality analysis that exploits the commonalities among the family and simultaneously accommodates variations among the members. The FAST process can be divided into two stages: domain engineering and application engineering. The first step is domain analysis in the domain engineering stage. Commonality analysis is the preferred method for this analysis, in which experts collect and document their knowledge of the SPL. To produce family products quickly and cheaply, an application engineering environment was developed based on a commonality analysis. In such an environment, domain-specific languages and architectural frameworks are often used. The second step in domain engineering is to translate the knowledge obtained from the commonality analysis phase into useful technology. This useful technology can include the creation of domain-specific and composable components. The application engineer uses the information obtained to produce the individual members of a product family. Feedback also plays a vital role in the modification of the environment. FAST focuses mainly on domain engineering and is used for the fast generation of individual products.

**Notation-Independent VM Approaches:** The SPLIT (Coriat et al., 2000) approach uses an extension of the UML for variability modelling. It uses a multilevel decision model to provide definitions of decision rules that are needed to identify the relations between decisions. This approach is dependent on an environment that uses UML, and intentionally, no particular notation is used to describe product-line assets to approach notation independence. Schmid and John (2004) also proposed a notation-independent representation for VM as a meta-model containing the following components: a decision model for variability effect characterization, a mechanism for the description of various decision interactions, a mechanism for variability resolution, a set of selector types, and a mapping of selector types to specific notations for expressing variation points in the artifacts. This approach complements Muthig's (2002) approach but focuses more on using product-line methods along with a comprehensive modelling approach for variability.

**Optimization-based VM Approaches:** Loesch and Ploedereder (2007) proposed an iterative semi-automated process and a variability optimization method for an SPL. It optimizes the number of required variable features without affecting the configuration of existing SPLs and future products. This approach also provides an interactive visualization of variability to help SPL engineers perform product derivation, variability management, and other tasks. The entire process consists of four phases: a) in variability documentation, a matrix is constructed relating product configurations and variable features to provide precise documentation of SPL variability; b) the variability prediction phase predicts the required future variability; c) in variability analysis, the matrix is used to analyze the use of current and future variability features in product configuration. Features are classified according to their usage, and constraints are identified; and d) in variability restructuring, the output of variability analysis is used to derive restructuring strategies to simplify variability and apply these strategies to SPL components. This approach provides limited support for the evolution of the variability.

**Variation point model (VPM) Approach:** VPM (Webber & Gomma, 2004) manages variation points from common requirements up to the design level. It models four types of variability: parameterization enables the user to define a population of attributes; inheritance makes it possible to choose variants from a limited set of choices; information hiding also enables a user to choose from a limited set of choices, but the interface is common, and the implementation is different, and the last type is evaluated using variation points. The VPM extends the definition of a variation point by including its variability mechanisms in addition to its mechanism. This approach describes the four views that are necessary for adequate communication of the variation points to the user. The four views are the requirements view, component variation point view, static variation point view, and dynamic variation point view. This approach models the variability of core asset component variation points and builds target system components from unique variants built from variation points. It provides more flexibility to the re-user and enables the creation and maintenance of unique variants.

**Configuration-based Approach**: Krueger's (2002) approach is based on configuration management. In SPL, the VM is a multidimensional problem of configuration management. The author used the "divide and conquer" approach to the variation management problem in software product lines. Variability management problems are divided into a collection of nine smaller problems and solutions. The proposed solutions to the nine problems are lightweight solutions that also help reduce the associated cost, risk, and time. The application of the proposed solution was also analyzed even with legacy software systems that have no proper documentation.

1)   Result of Comparing and contrasting phase

Research A focus group comprised three experts ( two from academia (SPLE researchers) and one from industry (dealing with VM projects). They compared the different phases of variability management against each other using the semantic comparison of the descriptions of each phase in the selected approaches, as shown in Table 3. The identified approaches were selected based on the inclusion criteria that mainly described variability management for an SPL. In the analytical technique step, the experts identified the twelve specific stages from the identified studies and numbered them in the leftmost column of Table 3, Appendix I.

Table 3 shows the different approaches that implicitly or explicitly discuss VM phases from the literature. The phases are divided into two domains: domain engineering and application engineering. Some approaches discuss phases from both domains, but some only focus on the application phase, such as Krueger (2002). After comparing and contrasting, a total of 12 phases were identified. Contrasts between these approaches, along with the metaphors and themes used, were identified throughout the process and are described below. Concepts, themes, and metaphors are italicized.

*Phase 1.* The approach taken by Kim et al. (2011) is exceptional in that it explicitly emphasizes business planning and product information as initial steps toward VM and calls them a scoping step that determines the boundaries of an SPL. FODA (Kang et al., 1990) and FORM (Kang, 1989) also help domain engineers in the identification of commonality and variability features in product lines. FAST (Ardis et al., 2000) is another approach that addresses the commonality analysis issue for SPLs. Junior et al. (2005) included a variability-tracing definition in the form of a variability-tracing model. The approach of Kim et al. (2011) consisted of both business and technical perspectives on the SPL for VM. This phase is important because it provides prerequisites for later stages.

*Phases 2 & 3.* Phase 2 involves the identification of variability. It seems that most of the approaches include a requirement view, C&V modelling, or a variation point view. Despite the different names used by the approaches, they all suggest that the organization must identify variability in SPLs as its first initiative toward VM. VPM (Webber & Gomaa, 2004) discussed the identification of requirements for SPLs from a technical perspective. Phase 3 includes the dependency view of variability in one product with other features, as discussed in the context of COVAMOF (Sinnema et al., 2004). These two phases can be merged because they highlight the importance of variability identification and their dependency on other SPL features. Itzik et al. (2016) introduced an approach to automate the requirements variability analysis based on ontological and semantic considerations. This approach analyzes and presents variability based on textual requirements only.

*Phases 4, 5, & 6.* The detailed phases 4, 5, and 6 are primarily related to the modelling of variability in most of the approaches that describe it differently. FAST (Ardis et al., 2000) refers to this phase as the creation of composable components, VPM (Webber & Gomaa, 2004) defines different views of static and dynamic variation, and Kumbang (2007) and Font et al. (2017) described this phase in the form of a model, but all these approaches contain the technical perspective of VM. The approach taken by Kim et al., (2011) is different because it describes the business service scenario along with architectural modelling from a business perspective.

*Phases 7, 8, & 9.* The next detailed phases deal with the actual implementation of variability along with the SPL application engineering phase. Junior et al. (2005) proposed a variability implementation mode and the tracing and control of variability from a technical perspective. The Krueger approach is based on configuration management and implements variability in the form of components. Kim et al. (2011) adopted a distinct approach in that they briefly discussed the variability in product design, analysis, and development. The FAST (Ardis et al., 2000) approach simply discussed variability under the generic architecture phase in a general manner. The Schmid and John (2004) approach described the decision model for variability and the identity of the actors and discussed the variability binding stage from both a business and technical perspective.

*Phases 10, 11, & 12.* The last detail phase is the optimization of variability, which is found in only a few approaches. The FAST (Ardis et al., 2000) approach briefly describes the testing phase for an SPL, but does not explain the testing phase for VM. Junior et al. (2005) attempted to describe the optimization of variability in the form of a configuration analysis of a particular product. Myllärniemi et al. (2016) also proposed theories for performance variability in software product lines and evaluated them using a case study. The most important approach in terms of optimization or evolution of variability is the Loesch and Ploedereder (2007) approach. It describes in detail how to analyze, predict, and restructure variability from both business and technical perspectives for better scalability and business portfolio management.

*Summary of comparison and contrast of features of VM approaches:*

A cross-modal comparison revealed two important themes. Figure 3 shows the results of the "comparison and contrasting stages" of the research methodology based on different identified phases. It also highlights the reciprocal relationship between the identified themes and phases from this stage. One theme is related to the business perspective of VM in SPLs, and its key concepts are business planning, business service scenarios, decision models of product variability, and portfolio management. Portfolio management refers to the management of different business services based on the implementation of variability. The other theme is related to the technology perspective of VM adoption in SPLs, and its identified key concepts are requirements identification, architectural modelling, implementation, and testing. Table 3 and Appendix I present the 12 detailed phases identified in the analysis. The first phase of business planning and product information is important from both perspectives. Phases 2 and 3 are merged because they cover the identification of SPL requirements from a technical perspective. Phases 4, 5, and 6 are clustered together because they address the variability modelling issue. Phases 7, 8, and 9 discuss the binding of variation points in terms of variability implementation. Phases 10, 11, and 12 deal with variability optimization from both technical and business perspectives for better scalability.

2) Result of the Reciprocal translation phase

It is about identifying underlying concepts and themes. The next step in the analysis is a reciprocal translation, which deals with the translation of studies into one another's terms.

At this stage, the surveyed studies are compared and contrasted with each other and with the dimensions identified in the comparison study. A reciprocal translation of the concepts and themes is presented in Table 4 in Appendix I and compared to the phases provided in the surveyed approaches. A tick mark against the concepts in a given set of cells represents the existence of that concept in the approach. First, the initial business planning phase seems to be related more to the business perspective but is also related to the technical perspective, whereas requirements identification is more related to the technical perspective. Business service scenarios and decision modelling both refer to the business perspective. Architectural modelling, implementation, and testing seem to be related more to a technical perspective.

*3.2 Result of Synthesis of Translation; Relating Concepts and Themes Revealing Underlying Metaphors.*

The synthesis of translation is the last step of the method and explores the underlying themes and concepts for metaphors. This step of the study confirmed the preliminary findings of previous reciprocal translations. The expert team sequenced and classified the initial concepts identified (such as business planning, requirements identification, business service scenarios, architectural modelling, decision model, implementation, and testing) with the two themes (business and technical perspectives). Business planning, business service scenarios, decision models, and portfolio management concepts were related to the business perspective, whereas requirements identification, architectural modelling, implementation, and testing concepts were related to the technical perspective. Next, all identified concepts were sequenced during the development and adoption phases. Finally, four phases/metaphors were identified across the two themes and eight concepts. Table 5 in Appendix I summarizes the definitions and corresponding phases of the metaphors. The four metaphors identified through this synthesis of the translation step are variability identification, variability modelling, variability implementation, and variability optimization.

*Variability identification metaphors:* The variability identification metaphor refers to the identification of variability requirements and the planning of product variability information. Requirement identification falls under the technical perspective, and business planning falls under the business perspective. This phase determines the SPL boundaries, and the variable-tracing model can be used to capture the variability identification requirements. An enhanced product map, feature list, or refined feature diagram can be the outcome of this metaphor. This metaphor can be positioned in the first phase of the VM adoption frame of reference as an initial step toward the VM process.

*Variability-modelling metaphor*: This metaphor consists of the variability modelling process, and most previous studies have addressed this issue. From a technical perspective, architectural modelling describes the variability that is allowed in products and model variations in a particular product. The business service scenario falls under the business perspective and models different business services from the standpoint of product variability. The outcomes of this metaphor can be refined using use case models or business service scenarios. A product feature map can also be constructed. This metaphor can be positioned after the identification of variability between business service scenarios and architectural modelling. The first two metaphors can be categorized as domain engineering.

*Variability implementation metaphors:* The variability-implementation metaphor consists of a decision model, product derivation, or implementation concepts. This metaphor can be aligned with these two concepts. This metaphor implements variability, and tracing and binding of different product variation points also occur in this phase. The variable product can be derived by binding the variation points. This metaphor falls under the implementation of the SPL stage and the decision model from a business perspective and also involves decisions about variability in the product line from a business perspective. The actual product configuration is a part of this metaphor.

**Variability optimization metaphor:** The final metaphor is variability optimization. Among the key concepts that can be aligned with this metaphor is the testing of variation point binding, and portfolio management from a business perspective deals with the management of this metaphor. The different services offered by a product also provide scalability. This metaphor can also involve a product configuration analysis. Addition, deletion, and updating of variation points can also be addressed using this metaphor which can be positioned at the last phase of VM, which is the adoption of a common SPL frame of reference.

The variability implementation metaphor can be categorized under application engineering.

## 4. Common Framework for VM of SPL

Several approaches have been identified from the literature and compared with each other for the meta-synthesis of qualitative analysis. Key concepts were extracted from these approaches because they used different descriptions. The underlying themes were identified by performing semantic analysis. Through reciprocal translation and synthesis of translation, underlying metaphors were extracted. Figure 4 shows the resulting common framework for VM adoption in the SPL for any organization.

The analysis of content identified two apparent themes: business and technical perspectives. For any organization to deliver competitive software products, these two themes are very important, and the present analysis distinguishes them from each other. In Figure 4, the technical and business themes are represented by the x- and y-axes, respectively. The relationships between phases and themes indicate four separate but interrelated metaphors: variability identification, variability modeling, variability implementation, and variability optimization. These metaphors are indicated on the diagonal of the figure as boxes with bold characters. Two themes are clearly shown by all the metaphors: technical and business perspectives.

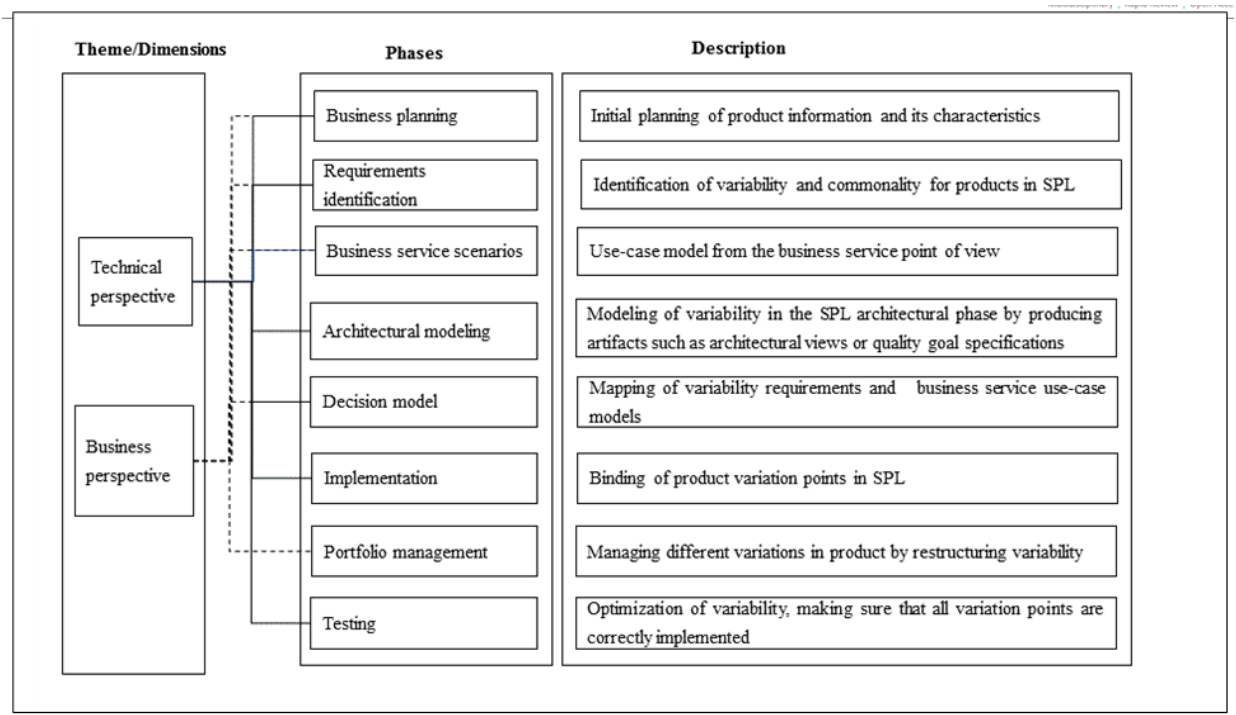

Figure 4. Relationship between themes and stages

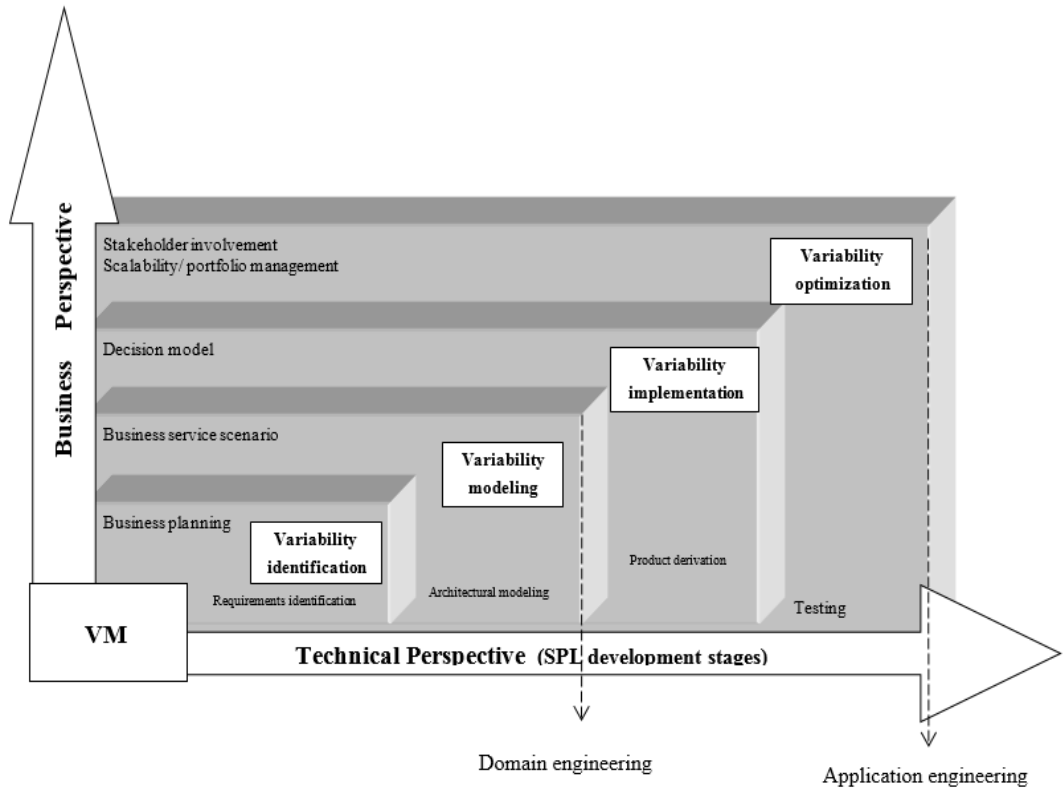

Figure 5. A common frame of reference for VM in SPL

The variability identification metaphor covers the concepts of requirements identification and business planning in parallel, whereas variability modeling covers architecture modeling and business service scenarios. Both metaphors fall under domain engineering. The variability implementation metaphor includes the concepts of a decision model and the implementation phase of an SPL, whereas variability optimization consists of testing and portfolio management.

Existing VM approaches seem to be fragmented in terms of their perspectives and applications, as revealed by reciprocal translation. None of the approaches in the literature are comprehensive enough that an organization can use it as a guideline for VM adoption. Management, organizational perspective, technology, and narrow focus are all present in a fragmented manner across various approaches. The use of a meta-synthesis qualitative analysis reveals a common frame of reference for VM adoption, which was presented in this study.

This common frame of reference is simple, yet comprehensive, and covers all the features provided by the various approaches. It can be used by any organization for VM adoption in SPLs. From a technological standpoint, the proposed frame of reference is an accumulative model for VM adoption. The first phase involves the identification of the business planning and variability requirements.

The second phase includes the modeling of product variability in a software product line and can be accomplished in many ways because most of the work related to VM is performed in this area. The third phase involves the actual implementation of variability features in a product, and the final phase deals with variability optimization to track variability features.

## 5. Evaluation of the Proposed Framework

To evaluate the common frame of reference, a qualitative study was performed. The case study approach was used to demonstrate the applicability of the common frame of reference. Generally, the case study approach is an appropriate strategy, when 'why' or 'how' questions are of primary interest (Yin, 2008). The goal of case studies is not to validate the proposed common frame of reference in this phase of the study, but to demonstrate its applicability in the SPLE industry. The study involved a set of semi-structured interviews to evaluate the proposed framework of VM practices in six organizations. We first describe the interview design, participants, and research methodology. We then present detailed interview results based on the interview data.

*5.1 Interview Design and Participant Data Collection*

We conducted semi-structured 36 interviews, among which some interviewers were employees of the same organization. The participants were involved in the product-line efforts of a single organization. The subjects involved in the study were two industrial companies that applied SPLE and variability modeling, two were SPLE consultant companies, one was a telecommunication infrastructure company, and the last was a mobile game development organization. Industrial companies are selected because they publish product-line efforts in terms of highly referenced experience reports.

The consultant companies were selected because they showed interest in our proposed framework and their employees' responses to our questionnaire. We invited participants to participate in the interview, and each interview lasted for almost one hour. A sample questionnaire is provided in Appendix II. The names of the respondents and their organizations were kept confidential for reasons of privacy. Participating organizations were informed that this is a research study and that, subsequently, neither the identity of the organization nor that of any individual would be disclosed in any publication.

The two consultancy companies were medium-sized organizations and developed custom software solutions for their clients. One is an IT consulting organization referred to as case study A. The respondents were IT consultants with 6-8 years of experience. The second is a web-based application organization referred to as Case Study B, and respondents of the interview are process managers and software architects with experience of 8-15 years working in a team of 5-7 members. The industrial organizations come under the category of large-scale and belong to automotive (referred to as case study C), and respondents are software architects with experience of 3-6 years that previously modelled and managed variability for the organization's product lines. The fourth category is the electro-mechanical producers' category (Case Study D), and respondents are software architects with a team size of 6-8 members having experience of 5-8 years. They developed the SPL for their main products. The fifth is a telecommunication large-scale organization, and its product line contains both software and hardware (case study E), and respondents are product line architects and managers having experience of 5-17 years. The mobile game organization is small and targets different mobile device hardware, software, and sales channel customization (case study F), and respondents are managers and software developers with experience varying from 5 to 15 years and working in a team size of 3-5 members.

*5.2 Interview Result*

Responses were collected from the respondents. We analyzed the collected data, and figure 5 presents a summary of the results.　　Figure 5 shows the activities performed in each phase of the presented framework for the VM in each case study.

*Case Study A & B:* Case study A is involved in an independent IT consulting company that offers expertise in product line research and involves selling methodologies and technologies for developing SPL. The respondents of the interviews were IT consultants. Case Study B was a web-based application development company. The respondents of the interviews were process managers and software architects. The developers in both organizations prefer to use the visual and configuration capabilities of the feature models. They don't follow the variability management in terms of the phases. They merged their variability identification and modeling phases. They only focus on the technical perspective of variability management.

*Case Study C & D:* The respondents were software architects that previously modelled and managed variability for the organization's product lines. Both case studies are similar in that they perform variability identification and variability, and modeling does not use any formal model. Stakeholders were involved only in the testing phase. They also do not follow any specific phases for variability management, as presented in the common frame of reference. Both case studies focused more on the technical perspective of variability management. They do not use specific decision models for the variation points.

*Case Study E:*　The respondents are product line architects and managers. They also merge two phases of the framework, that is, identification and modelling involve only a technical perspective. In the implementation phase, they did not follow any feature map. They developed their decision model and did not use any formal representation. The optimization phase does not involve all the stakeholders.

*Case Study F:* The respondents from this organization were software developers and managers. They only manage lightweight variability. More focused on a technical perspective

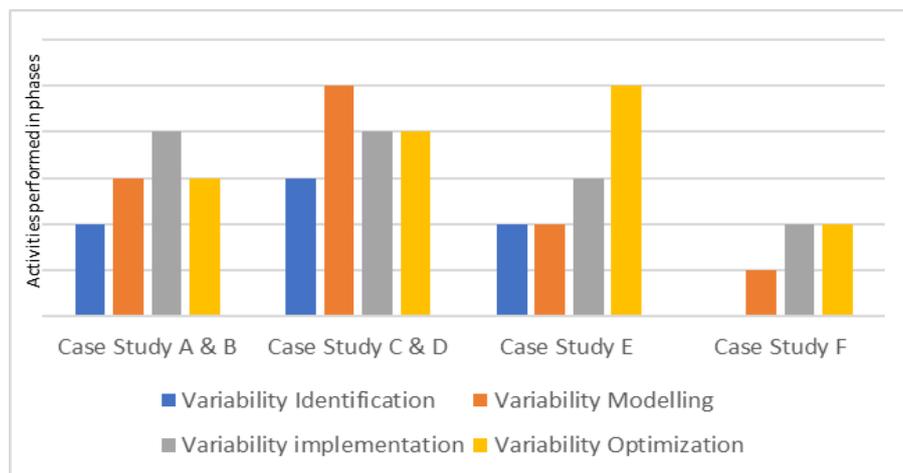

Figure 5. Framework evaluation qualitative study

In short, all the case studies are more focused on the technical perspective of the VM framework, as they cover most phases of the technical dimension presented in the framework, but pay very little attention to the business perspective of VM of SPL. They do not follow any specified variability framework for management. All the organizations found the presented framework and showed interest in the different metaphors of the phases. The results of this study solely depend on the respondents' perspective; therefore, it is subject to validity threats in an interview-based qualitative study. Another limitation is the number of respondents; we avoided any generalization to address this limitation.

**6. Conclusion and Future Directions**

Variability management from requirements identification to implementation is becoming a key business requirement. In this research, a qualitative meta-synthesis approach was used to contrast and compare various VM-related approaches found in the literature. A meta-synthesis approach was used to synthesize the results from the qualitative studies. A common frame of reference for VM was proposed as a result of this analysis. Putting metaphors in the context of the dimensions in which variability occurs and identifying key concepts provides a better understanding of variability management. This offers several opportunities for analysis and evaluation.

However, the proposed model is the first step toward understanding the VM. The authors evaluated the presented frame of reference by investigating the SPL organizations. The evaluation aims to identify the gap in practice by comparing it with the frame of reference. The results of the evaluation phase suggest that the organizations in practice only focus on one dimension. The presented frame of reference will help the organization to cover this gap in practice. In the future, we expect to work on elaborating metaphors from the two perspectives explored in this study and validate them through empirical research.

**Acknowledgments**

This study was supported by the RIF activity R17014 granted by Zayed University, Abu Dhabi, U.A.E.

**Appendix**

Table 1. Comparison of SPL approaches for variability management

| Approach | Underlying concept | Modeling Technique | Scope | Issues Addressed | Limitations |
|---|---|---|---|---|---|
| **FODA (Kang et al., 1990)** | Feature-oriented | Feature modeling | SPL life cycle | • Variability modeling<br>• Identification of commonality and variability | • Scope limited to SPLE<br>• Does not cover a particular VM |
| **FORM (Kan et al. 1989)** | Based on FODA | Feature modeling | •Requirements phase<br>• Design phase<br>•Implementation phase | Variability modeling | Limited scope: limited to modeling |
| **RequiLine (Von der Massen & Lichter, 2004)** | Based on FODA | Feature modeling | Requirements phase | Variability modeling Tool support | Limited scope: covers only one phase |
| **Ferber et al. (2002)** | Based on FODA | Views to represent dependency interaction | Requirements phase | How dependent variant interact | Limited to views and requirement phase |
| **Ye & Li (2005)** | Extension of Ferber's approach | View extension | Requirements phase | Variability modeling Evolution of variability | Limited scope: covers only one phase |
| **van der Hoek (2004)** | Component/ Koala-based | | Architecture-centric | Support any-time variability Binding time addressed | Limited scope to architecture phase |
| **Koalish (Asikainen et al. 2004)** | Component-based | | Architecture and configuration phases | Variability modeling Product derivation | Limited scope: covers only one phase |
| **Krueger (Krueger 2002)** | Configuration-based | N/A | Configuration phase | Product derivation File system level | Limited scope: covers only one phase |
| **COVAMOF (Junior et al. 2005; Sinnema et al., 2004)** | Based on the feature model | Variation point view and dependency view | Define the level of abstraction<br>a) features<br>b) architecture<br>c) component implementation | Variability modeling Product derivation | Limited scope: : covers only one phase |
| **Kumbang (Asikainen at al., 2007)** | FODA+ component (Koala)-based | | | | |
| **VPM (Webber & Gomaa, 2004)** | Variation point-based | UML extension | Requirements phase | Variability modeling | Limited scope: covers only one phase |
| **Muthig [68]** | Separation of variability representation from SPL artifacts + uses a decision model | Notation-independent | Requirements phase | Variability modeling | Limited scope: covers only one phase |
| **Schmid and John (2004)** | Customizability approach and use of decision model | Notation-independent | Requirements, architecture, and implementation phases | Variability modeling and emphasis on ease of adoption | |
| **FAST (Ardis et al. 2000)** | Process support | | | | No prescription of VM model |
| **DRM & MOON et al.(2005)** | Based on FODA | | Requirements phase | Identification of commonality and variability | Limited scope: covers only one phase |
| **Loesch and Ploedereder (2007)** | Optimization-based | | Maintenance phase | Evolution of variability | Limited to the maintenance phase |
| **Kim et al. (2011)** | Feature model | UML | Domain engineering & | VM in domain | Limited scope |

|  |  | extension | architecture |  | engineering | but comprehensive |

Table 2. Comparison of SPL Stages for Variability Management

| Approaches | SPL stages are covered by an approach | Mapping in comparison with SPL stages for Variability Management |
|---|---|---|
| **Muthig (2002)** | • Separation of variability from SPL artifacts<br>• Notation-independent | Requirements engineering phase |
| **COVAMOF (Junior et al. 2005; Sinnema et al., 2004)** | Modeling techniques<br>   a) Variation point view<br>     (i) variation point (ii) variant (iii) dependency<br>   b) Dependency view | Feature model<br>Architecture<br>Component implementation |
| **Koalish (Asikainen et al., 2004)** | Variability modeling | Modeling language for configuration-based approaches<br> 1) Components and compositional structure<br> 2) Connection points<br> 3) Attributes<br> 4) Constraints<br>Formalization (weight constraint rule language) - tool support |
| **Variation Management for SPLE Krueger (2002)** | Configuration-based approach—File system level<br>a) Basic configuration management<br>   i) version management ii) branch iii) baseline iv) branched baseline<br>b) Component composition<br>   i) composition management ii) branched composition management<br>   c) Software mass customization<br>   i) Variation point management, ii) customization management iii) customization composition management | SPLE artifacts under variation management<br> 1) Domain engineering- common & variant artifacts.<br> 2) Product instantiation- instantiation infrastructure<br> 3) Product development – product instances<br>Use-product instances |
| **KobrA (Atkinson et al., 2000)** | Integration of product-line concept into component-based development; VM not discussed in particular | 1) Framework engineering phase<br>   i) context realization, ii) component specification, iii) component realization<br>2) Application engineering<br>   i) application context realization, ii) framework instantiation |
| **Schmid and John (2004)** | Limited to component-based development only-<br>   a) decision model<br>   b) interactions<br>   c) relations<br>   d) variation types<br>   e) specific mapping | Traceability of variability in all kinds of SPL lifecycle artifacts, both horizontally and vertically |
| **Kim et al. (2011)** | a) Domain engineering<br>  i) Scoping: business planning and product information<br>  ii) C&V modeling – Control feature list, enhanced product map, refined feature diagram<br>  iii) Architecture modeling- Inputs are refined use case model, business service case scenario, and outputs are quality goals specification, architecture view<br>b) Application engineering<br>  i) product analysis<br>  ii) product design<br>    product development | Mainly focused on domain engineering in SPL |
| **FAST (Ardis et al. 2000)** | a) Commonality analysis<br>Creation of domain-specific language & creation of composable component | |
| **Loesch and Ploedereder (2007)** | Describes participants like core asset developers, product developers, technology experts, marketing managers, product line manager<br>Phases for optimization process:<br>  a) variability documentation<br>  b) variability prediction<br>  c) variability analysis<br>  d) variability restructuring | This approach is dependent on evolving SPL with existing variability for the optimization process. |
| **FORM (Kang et al. 1989)** | VM is not explicitly discussed | 1) Domain engineering<br>   i) Domain analysis & feature |

| | | | |
|---|---|---|---|
| | | | modeling<br>　　ii) Architectural & component modeling<br>2) Application engineering<br>　　i) Requirements analysis & feature selection<br>　　ii) Architectural model selection & application development. |
| **Junior et al. (2005)** | a) Variability tracing definition- variability tracing model<br>b) Variability identification<br>c) Variability delimitation<br>d) Identification of mechanisms for variability management. | | 5) Requirements identification (UML-based) |
| **VPM (Webber & Gomaa, 2004)** | a) Requirements view<br>b) Component variation point view<br>c) Static variation point view<br>d) Dynamic variation point view | | Requirements to design level. |

Table 3. VM Phases Identified From Approaches

| Authors | FAST (Ardis et al. 2000) | Krueger (2002) | Schmid and John (2004) | VPM (Webber & Gomaa, 2004) | Junior et al. (2005) | COVAMOF (Sinnema et al. 2004) | Kumbang (Asikainen et al. 2007) | Loesch and Ploedereder (2007) | Kim et al. (2011) |
|---|---|---|---|---|---|---|---|---|---|
| Year | 2000 | 2002 | 2004 | 2004 | 2005 | 2004-2008 | 2007 | 2007 | 2011 |
| Phase # | 5 | 3 | 5 | 4 | 4 | 2 | 3 | 4 | 3 |
| **Domain Engineering** | | | | | | | | | |
| 1 | Commonality analysis | | | | Variability tracing definition | | | | Scoping |
| 2 | Creation of domain-specific language for variability identification | | | Requirements view | Variability identification | Variation point view | Meta layer | Variability documentation | C & V modeling |
| 3 | | | | | | Dependency view | | | |
| 4 | Composable component creation | | | Component variation view | Variability delimitation | | Kumbang model | | Architectural modeling |
| 5 | | | | Static variation point view | | | Instance layer | | |
| 6 | | | | Dynamic variation point view | | | | | |
| **Application Engineering** | | | | | | | | | |
| 7 | Generic architecture | Basic configuration mgmt. | Decision model | | Variability implementation mode | | | | Product analysis |
| 8 | | Component composition | Mapping | | Variability tracing & control | | | | Product design |
| 9 | | Software mass customization | | | | | | | Product development |
| 10 | Testing phase | | | | Configuration analysis of specific product | | | Variability prediction | |
| 11 | | | | | | | | Variability analysis | |

| 12 | | | | | | | | Variability restructure-ing | |

Table 4. Underlying Metaphors and Themes of the VM Phase Model

| Metaphor | Description | Stages/Concepts | |
|---|---|---|---|
| | | Business Perspective | Technical Perspective |
| Variability identification | Identification of product variability for SPL | Business planning | Requirements identification |
| Variability modeling | Modeling of variation points for specific products | Business service scenarios | Architectural modeling |
| Variability implementation | Product derivation and binding of version points | Decision model | Implementation |
| Variability optimization | Restructuring of variability, addition, deletion, and updating of variation points | Portfolio management | Testing |

Table 5. Metaphors: their definition, related stages, and themes

| Metaphor | Themes | Concepts | FAST (Ardis et al. 2000) | Krueger (2002) | Schmid and John (2004) | VPM (Webber & Gomaa, 2004) | Junior et al. (2005) | COVAMOF (Sinnema et al. 2004) | Kumbang (Asikainen et al. 2007) | Loesch and Ploededer (2007) | Kim et al. (2011) |
|---|---|---|---|---|---|---|---|---|---|---|---|
| Variability identification | Business/technical perspective | Product information / requirements identification | ✓ | | | ✓ | ✓ | | ✓ | | ✓ |
| Variability modeling | Business/ technical perspective | Business services scenarios/ architectural modeling | ✓ | | ✓ | ✓ | ✓ | ✓ | ✓ | ✓ | ✓ |
| Variability implementation | Business/ technical perspective | Decision model/ implementation | | ✓ | ✓ | | ✓ | | ✓ | ✓ | ✓ |
| Variability optimization | Business/ technical perspective | Portfolio mgmt./testing scalability | | | | | | | | ✓ | |

**Appendix B**

Table 6. Sample semi structured Interview questionnaire

| Variability Management Framework | |
|---|---|
| **Domain Engineering** | |
| **Variability Identification (Requirement identification & Business planning)** | 1. Is there a formal/informal process to identify variability in SPLE? |
| | 2. Is enhanced product map developed during identification phase or not? |
| | 3. Is feature list identified or not? |
| | 4. The alignment of business objective is performed at the identification phase or not? |
| | 5. All stakeholders are involved at this stage or not? |
| **Variability Modeling (Architectural Modelling & Business service scenario)** | 1. Is Meta-Model developed formally for variation points? |
| | 2. Is formal representation used such as UML/Arbitrary modeling? |
| | 3. Are models used for the different business scenarios from the standpoint of product variability? |
| | 4. Are product feature maps used for modeling variation points in SPLE? |
| | 5. Are architectural components are usually derived from requirement mapping mechanisms? |
| **Application Engineering** | |
| **Variability implementation (Product derivation & Decision model)** | 1. Is decision modelling used for variability implementation and aligned with business objective? |
| | 2. Are product variation points are traced and bind bases on decision model or they variate? |
| | 3. Does your project team apply developed feature map fully for product variation points? |
| **Variability Optimization (Testing & Scalability, Portfolio Management)** | 1. After binding take place, team always perform testing of variation points or not? |
| | 2. Is portfolio management is used for SPL? |
| | 3. Is product configuration analysis is performed or not? |
| | 4. Did you perform any update for variation points of product based on configuration analysis? |
| | 5. All Stakeholders are involved in this phase or not? |